\documentclass[superscriptaddress,twocolumn,showpacs,prb]{revtex4-1}
\usepackage[utf8]{inputenc} 
\usepackage{amsmath}
\usepackage{braket}
\usepackage{graphicx}
\usepackage{amsfonts}
\usepackage{pgfplots}
\usepackage{csquotes}
\usepackage{hhline}
\usepackage{amssymb}
\usepackage{listings}
\usepackage{color}

\definecolor{codegreen}{rgb}{0,0.6,0}
\definecolor{codegray}{rgb}{0.5,0.5,0.5}
\definecolor{codepurple}{rgb}{0.58,0,0.82}
\definecolor{backcolour}{rgb}{0.95,0.95,0.92}
 
\lstdefinestyle{mystyle}{
    backgroundcolor=\color{backcolour},   
    commentstyle=\color{codegreen},
    keywordstyle=\color{magenta},
    numberstyle=\tiny\color{codegray},
    stringstyle=\color{codepurple},
    basicstyle=\footnotesize,
    breakatwhitespace=false,         
    breaklines=true,                 
    captionpos=b,                    
    keepspaces=true,                 
    numbers=left,                    
    numbersep=5pt,                  
    showspaces=false,                
    showstringspaces=false,
    showtabs=false,                  
    tabsize=2
}

\usepackage{subfigure}

\usepackage{dcolumn}
\usepackage{tabularx}
\usepackage[colorlinks=true,linkcolor=blue,citecolor=blue,urlcolor=blue]{hyperref}
\usepackage{longtable}
\usepackage{braket}
\usepackage{float}
\newcolumntype{C}{>{\centering\arraybackslash}X}
\begin{document}
\title{Demonstration of Minisuperspace Quantum Cosmology Using Quantum Computational Algorithms on IBM Quantum Computer}

\author{Anirban Ganguly}
\email{anirbangangulynostalgic@gmail.com}
\affiliation{Physics Department, \\Presidency University, Kolkata 700073, West Bengal, India}

\author{Bikash K. Behera}
\email{bkb18rs025@iiserkol.ac.in}
\author{Prasanta K. Panigrahi}
\email{pprasanta@iiserkol.ac.in}
\affiliation{Department of Physical Sciences,\\ Indian Institute of Science Education and Research Kolkata, Mohanpur 741246, West Bengal, India}

\begin{abstract}
Quantum computers promise to efficiently solve important problems that are intractable on a conventional computer. Quantum computational algorithms have the potential to be an exciting new way of studying quantum cosmology. In quantum cosmology, we learn about the dynamics of the universe without constructing a complete theory of quantum gravity. Since the universal wavefunction exists in an infinite-dimensional superspace over all possible 3D metrics and modes of matter configurations, we take minisuperspaces for our work by constraining the degrees of freedom to particular 3D metrics and uniform scalar field configurations. Here, we consider a wide variety of cosmological models. We begin by analyzing an anisotropic universe with cosmological constant and classical radiation. We then study the results for higher derivatives, Kaluza-Klein theories and string dilaton in quantum cosmology. We use IBM's Quantum Information Science Kit (QISKit) python library and the Variational Quantum Eigensolver (VQE) algorithm for studying these systems. The VQE algorithm is a hybrid algorithm that uses the variational approach and interleaves quantum and classical computations in order to find the minimum eigenvalue of the Hamiltonian for a given system.
\end{abstract}

\maketitle

\section{Introduction}
Predictions of Superstrings, Supergravity and higher-dimensional Kaluza-Klein theories, which try to unify the four fundamental interactions of nature are for the present, as well as for the near future, well beyond the reach of our particle accelerators \cite{qgc_SolomonsUCT1994}. In quantum cosmology, we attempt to describe a quantum state of the universe, without the actual unification into a ``Theory of Everything". The Einstein-Hilbert action for a particular cosmological model is quantized by the method of canonical quantization. The Wheeler-DeWitt constraint is applied to the obtained Hamiltonian and the equation is solved for the universal wavefunction \cite{qgc_HalliwellWS1991}.

Supercomputers are huge technological breakthroughs of the modern era. However,understanding how atoms, and molecules, interact with each other on a larger scale is beyond the scope of such technologies. Quantum systems are too mathematically complex for today’s limited computing power. For example,it is impossible without approximations to calculate the properties of atoms, which are even slightly more complicated than the simple hydrogen atom.

Here, quantum simulation can provide a solution.Richard Feynman \cite{qgc_FeynmanIJTP1982} introduced the core idea of quantum simulations as long back as in 1982.Intractable quantum problems can be solved by simulating \cite{qgc_SommaQIC2016} them with other controllable quantum systems that resemble them. It is done by running an experiment, called a quantum simulator which yields a description of the system being studied.Thus, quantum simulations are ahead of the most powerful supercomputers and will enable the development of new tools at the forefront of research. For example,its applications in high-temperature superconductivity, would be beneficial for tackling climate change. Quantum simulation can also have wide-ranging applications in the field of quantum cosmology, the subject of our paper. We use an open-source development kit, ``QISKit", founded by IBM \cite{qgc_SantosRBEF2017}. QISKit provides tools for various tasks such as creating quantum circuits, performing optimizations, simulations, and computations. The QISKit Aqua library contains an implementation of the VQE algorithm which is the most suited for our purpose. It also provides the classical exact eigensolver algorithm which we use to compare our results. In this paper, we use the VQE algorithm to apply the Wheeler De Witt equation on four different quantum cosmological models, each with its own significance. We use two qubits for simulating each dimension of a system. 

IBM has become a huge platform for researchers in the field of quantum computation and quantum information \cite{qgc_IBM}, where users are allowed to run quantum computational algorithms on the 5-qubit (ibmqx2, ibmqx4), 16-qubit quantum processors (ibmqx5), and one 32-qubit simulator (IBM qasm simulator). A large number of quantum computing researches such as quantum simulational works \cite{qgc_ZhukovQIP2018,qgc_MalikRG2019,qgc_MalikarXiv2019,qgc_SchuldEPL2017,qgc_ManabputraarXiv2018,qgc_ViyuelanpjQI2018}, developing quantum error correcting algorithms \cite{qgc_GhoshQIP2018,qgc_Roffe2018,qgc_SatyajitQIP2018,qgc_HarperarXiv2018,qgc_SingharXiv2018,qgc_WarkeRG2019}, testing quantum algorithms \cite{qgc_BabukhinarXiv2019,qgc_GarciaJAMP2018,qgc_GangopadhyayQIP2018,qgc_YalcinkayaPRA2017,qgc_quantumsecretsharing,qgc_BeheraQIP2017,qgc_Solano2arXiv2017,qgc_BeheraQIP2019,qgc_RajiuddinRG2019,qgc_ChatterjeeRG2019,qgc_DashRG2019,qgc_DuttaarXiv2018}, developing quantum applications \cite{qgc_BeheraQIP12019,qgc_PalRG2018,qgc_SinghRG2019,qgc_MahantiQIP2019,qgc_MishraRG2019} have already been performed.

The rest of the paper is organized as follows: In Section \ref{qgc_Sec2}, we discuss briefly over the method of canonical quantization of the Einstein-Hilbert action of a universe and the formulation of the Wheeler De Witt equation for a minisuperspace. In Section \ref{qgc_Sec3}, we provide an introduction to quantum computation. In Section \ref{qgc_Sec4}, we discuss the discretization method used for the construction of the Hamiltonians corresponding to different models. In Section \ref{qgc_Sec5}, we explain the variational approach to quantum mechanics and the variational quantum eigensolver (VQE) algorithm. In Section \ref{qgc_Sec6}, we verify the Wheeler De Witt equation for different quantum cosmological models using the tools that we have developed. In Section \ref{qgc_Sec7}, we establish our results and show the optimized quantum circuits representing our solved universal wavefunction. In Section \ref{qgc_Sec8}, we conclude by suggesting some future prospects of our research.

\section{The Wheeler De Witt Equation and Minisuperspaces \label{qgc_Sec2}}
The Einstein-Hilbert action that results in the Einstein field equations in vacuum, through the principle of least action is given by

\begin{equation}
\label{qgc_Eq1}
S_{EH}=\frac{c^4}{16\pi G}\int d^4 x (-g)^{\frac{1}{2}} R  
\end{equation}

where R is the Ricci scalar, g is the determinant of the metric tensor matrix, c is the speed of light and G is the universal gravitational constant.
Total action of the world will be accompanied by Langrangian terms to account for matter fields \cite{qgc_ZeePUP2010} as well as the cosmological constant.

\begin{equation}\label{qgc_Eq2}
S=\int[\frac{c^4}{16\pi G}(R-2\Lambda)+L_M](-g)^{\frac{1}{2}}d^4x
\end{equation}

Other matter or radiation components can also be added to obtain the total action. The topology of our universe is taken as a 3-surface with metric $h_{ij}$ where i,j=1,2,3 with some matter field configuration $\phi$, embedded in a 4-D manifold on which 4-metric is $g_{\mu\nu}$ where $\mu,\nu=0,1,2,3$ is given by

\begin{equation}
\label{qgc_Eq3}
ds^2=g_{\mu\nu}dx^{\mu}dx^{\nu}=-(N^2-N_iN^i)dt^2+2N_idx^idt+h_{ij}dx^idx^j
\end{equation}

where N=lapse(arbitrary), $N_i$=shift. The Wheeler De Witt equation \cite{qgc_ShestakovaIJMPD2018} $H\psi=0$ is a constraint on the Hamiltonian H derived from the action in Eq. \eqref{qgc_Eq1} and $\psi$ represents the universal wavefunction that exists in an infinite-dimensional space of all the possible 3-metrics $h_{ij}(x)$ and matter field configurations $\phi(x)$, called the superspace \cite{qgc_HalliwellWS1991}. To make the problem solvable, we need to restrict the degrees of freedom of the universe to a finite number, giving a finite dimensional space called minisuperspace. Usually, in minisuperspace, the lapse is taken to be homogeneous and the shift is taken as zero. Hence, now, from Eq. \eqref{qgc_Eq2},

\begin{equation}
\label{qgc_Eq4}
ds^2=-N^2(t)dt^2+h_{ij} dx^i dx^j
\end{equation}

the 3-surface metric $h_{ij}$ is taken as homogeneous so that it can now be described by a finite number of functions of t(histories): $q^\alpha(t)$ where $\alpha$=0,1,2...n-1. The simplest case is the homogeneous and isotropic Friedmann-Robertson-Walker minisuperspace where $h_{ij}dx^idx^j=a^2(t)d\Omega_3^2$, $d\Omega_3^2$ is the metric on the 3-sphere and $q^\alpha(t)=a(t)$. We deal with other more complicated minisuperspaces in our paper as presented in Section \ref{qgc_Sec5}.

\section{Quantum Computation \label{qgc_Sec3}}

Quantum computation \cite{qgc_NielsenCUP2010} uses the properties of quantum mechanics: superposition and entanglement to encode information in units called the qubits which are fundamentally different from their classical equivalent, bits (0 and 1). Qubits are the 2-state quantum systems which can be physically represented through electron spins and photon polarizations while in the Hilbert space they can be described as,

\begin{equation}
\label{qgc_eq5}
\Ket{b}=\alpha_1\Ket{0}+\alpha_2\Ket{1}=\begin{bmatrix}
\alpha_1\\
\alpha_2\\
\end{bmatrix}
where ||\alpha_1|^2+||\alpha_2|^2|=1
\end{equation}

\begin{figure}[H]
\centering
\includegraphics[scale=0.7]{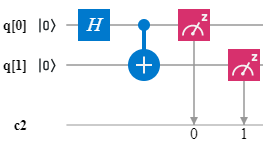}
\caption{Quantum circuit generating the state, $\Ket{\Psi}=\frac{|00\rangle+11\rangle}{\sqrt{2}}$.}
\label{qnm_Fig1}
\end{figure}

In a quantum circuit, the qubits are acted on by quantum gates, which can be described in Hilbert space as unitary matrices. Single-qubit gates include rotations of the qubits about the x-, y- and z-axes, the Pauli matrices, and the Hadamard gate which can put a qubit state into an equal superposition state. The most important, two-qubit gate is the controlled-not or ``CNOT" gate, which is commonly used to entangle two neighboring qubits. If we begin with n qubits in our quantum circuit, we obtain a state that is a superposition of $2^n$ basis states. So, when we make our measurements, we prepare an ensemble to get a probabilistic distribution of the results. A simple quantum circuit illustrating both superposition and entanglement is shown in Fig. \ref{qnm_Fig1}.

\section{Discretization of the Hamiltonians \label{qgc_Sec4}}
For the computation of any Hamiltonian to be possible, we need to work with a finite number of elements so instead of having a continuous eigen spectrum of position we need to discretize the space \cite{qgc_JainRG2019}. If we discretize a 2-dimensional space with x,y$\epsilon$[-L,L] such that we have N eigenvalues for each of x and y, then we obtain a mesh of $N^2$ spatial elements. Each element corresponds to a specific eigenvalue for x and y. If our mesh is centered at [0,0], our position operator will be an N*N matrix with the position eigenvalues as the diagonal elements. For 2 qubits and N=4,

\begin{eqnarray}
   x^d = \begin{bmatrix}
   -2&0&0&0 \\
   0&-1&0&0 \\
   0&0&0&0 \\
   0&0&0&1\\
   \end{bmatrix}
   \label{qnm_Eqn15}
\end{eqnarray}

\begin{figure}[H]
\centering
\includegraphics[scale=0.3]{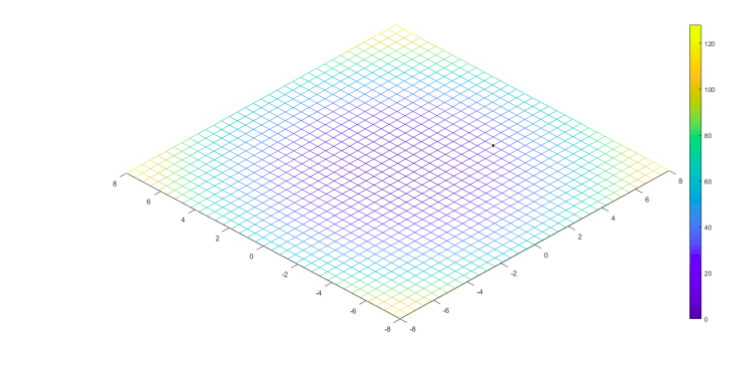}
\caption{Discrete two dimensional space. The colour bar indicates the amount of stretch required to map these points to harmonic oscillator potential \cite{qgc_JainRG2019}.}
\label{qgc_Fig1}
\end{figure}

The momentum operator can then be found by taking a discrete quantum fourier transform \cite{qgc_QFTWiki} of the position operator and an inverse discrete quantum fourier transform subsequently:

\begin{equation}
p^d=(F^d)^{-1} x^d F^d
\end{equation}

where $F^d$ stands for the discrete quantum fourier transform matrix whose each element is given by:

\begin{equation}
[F^d]_{j,k}=\frac{exp(i2\pi jk/N)}{N^{1/2}}
\end{equation}

Having obtained both the position and momentum operators, we can use them to find the discrete form of our Hamiltonian. For a quantum harmonic oscillator of one dimension, the Hamiltonian will be:

\begin{equation}
\label{qgc_Eq9}
H^d=\frac{(x^d)^2}{2}+\frac{(p^d)^2}{2}
\end{equation}

Since the Pauli basis ($X,Y,Z,I$) is a complete basis, to implement in a quantum circuit, we can decompose the discretized Hamiltonian into Pauli gates. The Hamiltonian for one-dimensional quantum harmonic oscillator in Eq. \eqref{qgc_Eq9} can be decomposed as 

\begin{equation}
\label{qgc_Eq10}
\begin{split}
H^d=0.589286 & ( I \otimes I )\\
-0.196429 & ( I \otimes Y )\\
+0.0982143 & ( I \otimes Z )\\
-0.0982143 & ( X \otimes I )\\
+0.196429 & ( X \otimes Y )\\
+0.196429 & ( Z \otimes I )\\
+0.196429 & ( Z \otimes Z )
\end{split}
\end{equation}

For higher dimensional models, we operate our Pauli gates on different spaces corresponding to different dimensions. The decomposed Hamiltonian for a 2-dimensional quantum harmonic oscillator will then be $(H^d\otimes II)+(II \otimes H^d)$ which from Eq. \eqref{qgc_Eq10} can be written as:

\begin{equation}
\begin{split}
1.178572 & ( I \otimes I\otimes I \otimes I )\\
-0.196429 & ( I \otimes Y\otimes I \otimes I )\\
+0.0982143 & ( I \otimes Z\otimes I \otimes I )\\
-0.0982143 & ( X \otimes I\otimes I \otimes I )\\
+0.196429 & ( X \otimes Y\otimes I \otimes I )\\
+0.196429 & ( Z \otimes I\otimes I \otimes I )\\
+0.196429 & ( Z \otimes Z\otimes I \otimes I )\\
-0.196429 & ( I \otimes I\otimes I \otimes Y )\\
+0.0982143 & ( I \otimes I\otimes I \otimes Z )\\
-0.0982143 & ( I \otimes I\otimes X \otimes I )\\
+0.196429 & ( I \otimes I\otimes X \otimes Y )\\
+0.196429 & ( I \otimes I\otimes Z \otimes I )\\
+0.196429 & ( I \otimes I\otimes Z \otimes Z )
\end{split}
\end{equation}

In this way, we can simulate any N-dimensional Hamiltonian into a quantum circuit. In our work, we will utilize this method to deal with 2-dimensional cosmological models.

\section{The Variational Quantum Eigensolver Algorithm \label{qgc_Sec5}}
Any normalized wavefunction can be written as a superposition of its eigenstates:

\begin{equation}
\label{qnm_Eq6} \psi=\sum c_n \psi_n
\end{equation} 

such that $H\psi_n=E_n \psi_n$ and $1=\big<\psi|\psi\big>=\sum|c_n|^2$. Now, since the ground state energy $E_g\leq E_n$, the expectation value
can be written as,

\begin{equation}\label{Eqn7}
\big<H\big> =\big<\psi|H|\psi\big>=\sum E_n|c_n|^2\geq E_g\sum|c_n|^2=E_g
\end{equation}

Thus, we obtain the variational principle of quantum mechanics \cite{qgc_GriffithsPPH2004}: $\big<H\big>\geq E_g$. In the VQE algorithm, we can take any trial wavefunction $\psi$ in the form of a quantum circuit. For this, we use the Ry variational form provided in the QISKit library. Then for obtaining the tightest bound on our ground state, we have to minimize the expectation $\big<H\big>$ with respect to the parameters of our circuit. This minimization is carried out through a classical optimization method called the simultaneous perturbation stochastic approximation (SPSA) \cite{qgc_SpallIEEE1992}. Once the quantum-classical hybrid algorithm (VQE) \cite{qgc_PeruzzoPRD2014} is carried out, we get the optimized variational form and the upper bound for the ground state energy.

\section{Quantum Cosmological Models}\label{qgc_Sec6}
\subsection{Bianchi-Type IX universe with $\Lambda,\gamma$ and small anisotropies}

Most of the works in quantum cosmology have been around homogeneous and isotropic Friedmann-Robertson-Walker minisuperspace models with topology $S^3$. However, considering anisotropic universes may allow us to obtain a more realistic insight into the fate of our universe. A Kantowski-Sachs minisuperspace \cite{qgc_LaflammePRD14}, for example, is an anisotropic universe with an extra anisotropic degree of freedom and topology $S^1\times S^2$. The Bianchi-Type IX minisuperspace \cite{qgc_AmsterdamskiPRD1985}, on the other hand, has two anisotropic degrees of freedom. It is given by the metric:

\begin{equation}
ds^2=-\sigma^2N^2(t)dt^2+\frac{\sigma^2}{4}a^2(t)(e^{2\beta(t)})_{ij}\sigma^i\sigma^j
\end{equation}

where

\begin{equation}
\begin{split}
\sigma^1&=cos\psi d\theta+sin\psi sin\theta d\phi\\
\sigma^2&=sin\psi d\theta -cos\psi sin\phi d\phi\\
\sigma^3&=d\psi+cos\theta d\phi
\end{split}   
\end{equation}

Constant $\sigma^2=\frac{l^2}{24\pi^2}$ and N(t) is a lapse function. $\beta$ is a time independent traceless diagonal matrix. We use the parametrization:

\begin{equation}
\begin{split}
    \beta_{11}&=\beta_+ +\sqrt{3}\beta_-\\
    \beta_{22}&=\beta_+-\sqrt{3}\beta_-\\
    \beta_{33}&=-2\beta_+
\end{split}    
\end{equation}

As explained in Section \ref{qgc_Sec2}, we can obtain the Einstein-Hilbert action for the minisuperspace. On adding components for the cosmological constant $\Lambda$ and $\gamma$ in the action, the Wheeler De Witt equation $H\psi=0$ takes the form:

\begin{equation}
   \Bigg[\frac{\partial^2}{\partial a^2}-\frac{1}{a^2}\bigg[\frac{\partial^2}{\partial\beta_+^2}+\frac{\partial^2}{\partial\beta_-^2}\bigg]+(\Lambda a^4-a^2+\gamma)+8a^2(\beta_+^2+\beta_-^2)\Bigg]\psi=0
\end{equation}

For small anisotropies $\beta_+ and \beta_-$, the equation depends only on $\beta=(\beta_+^2+\beta_-^2)$ so we can restrict it to a single anisotropic degree of freedom and we get the following equation:

\begin{equation}
   \Bigg[\frac{\partial^2}{\delta a^2}-\frac{1}{a^2}\frac{\partial^2}{\partial\beta^2}+(\Lambda a^4-a^2+\gamma)+8a^2\beta^2\Bigg]\psi=0
\end{equation}

The 2-dimensional Hamiltonian for our system is:

\begin{equation}
   H_1=-p_a^2+\frac{p_\beta^2}{a^2}+(\Lambda a^4-a^2+\gamma)+8a^2\beta^2
\end{equation}

We can now use the VQE algorithm to verify the Wheeler De Witt constraint on this Hamiltonian. Our resulting ground state eigenvalue is $0.4829572576220993\pm 0.12032107873580308$ for classical radiation density constant $\gamma=0.99$ and cosmological constant $\Lambda=2.480000000000011$. The classical exact eigensolver algorithm gives the result as $2.0707928270104423\times10^{-5}$. Fig. \ref{qgc_Fig3} shows the optimized quantum circuit representing our resulting solution for the universal wavefunction.

\subsection{Higher Derivative Cosmology}
For isotropic and anisotropic universes alike, it has been found in general that it is necessary to include a massive scalar field \cite{PedramPLB2009} in order to yield wavefunctions that on classical approximation, gives Lorentzian solutions of the field equations, which have a period of exponential expansion and approach isotropy at the present epoch, thus agreeing with observations. However, it is possible to avoid the inclusion of this massive scalar field term by resorting to quantum corrections to the gravitational action. These perturbations will lead to an effective gravitational action containing higher order curvature terms \cite{qgc_HawkingNP1984}. We take the action to be of the form:

\begin{equation}
\begin{split}
     S&=\frac{-1}{16\pi}\int d^4x \sqrt{g}(R-\alpha C^{\mu\nu\rho\sigma}C_{\mu\nu\rho\sigma}+\beta R^2)\\
    &-\frac{1}{8\pi}\int d^3x\sqrt{h}\big(K(1+2\beta R)-4\alpha K_{ij} C^{\mu i \nu  j}\eta_{\mu}\eta_\nu\big) 
\label{qgc_Eq20}
\end{split}    
\end{equation}

where $C_{\mu\nu\rho\sigma}$ is the Weyl Tensor and $\eta^\alpha$ is the unit normal to the boundary. The simplest minisuperspace model i.e. a homogeneous and isotropic universe is taken so the C-squared term is zero and can be neglected. The $R^2$ term on the other hand, behaves like a massive scalar field. The metric taken is of the form:

\begin{equation}
    ds^2=\frac{4a^2}{3\pi}(d\eta^2+d\omega_3^2)
\end{equation}
where

\begin{equation}
    d\eta=\frac{dt}{a}
\end{equation}

The action from Eq. \eqref{qgc_Eq20} is then:

\begin{equation}
    S=-\int d\eta\big(\frac{1}{a^2}-\frac{a''}{a^3}+\frac{9}{2}\pi\beta(\frac{1}{a^2}-\frac{a''}{a^3})\big)a^4
\end{equation}

Taking $Q=a(1+2\beta R)$ and x=Q-a,y=Q+a for convenience, our Hamiltonian

\begin{equation}
    H_2=-\frac{p_x^2}{4}+\frac{p_y^2}{4}-\frac{(y^2-x^2)}{4}+\frac{x^2}{8\widetilde{\beta}}(x-y)^2
\end{equation}

The VQE algorithm for $\widetilde{\beta}=0.042808219$ yields ground state energy eigenvalue $-0.07876069717108591\pm1.3971389629283328$. On applying the classical exact eigensolver algorithm, we get the eigenvalue as -1.6494040350599992. Fig. \ref{qgc_Fig4} shows the optimized quantum circuit generating the state vector.

\subsection{String Dilaton Cosmology}
String theory is one of the foremost proposals of unification into a theory of everything. It is relevant to cosmology as it can be investigated at Planck scale. We take the tree level dilaton-graviton string-effective action \cite{qgc_CapozzielloIJMPD1993} for a 4-dimensional spacetime as 

\begin{equation}
S=\int d^4x\sqrt{-g}e^{-2\phi}[R+\Lambda+4(\partial\phi)^2+..]
\end{equation}

where $\phi$ stands for the dilaton field. The dots indicate other antisymmetric tensor terms and scalar matter fields that we ignore. The minisuperpace is taken as a homogeneous and isotropic Friedman-Robertson-Walker metric. Using the transformation $2\Phi=2\phi-3\ln{a}$ and taking $z=\ln{a}$, we get the Hamiltonian for the system as

\begin{equation}
H_3=-\frac{p_z^2}{12}+\frac{p_\Phi^2}{16}+2\Lambda e^{-4\Phi}
\end{equation}

The result of using the VQE algorithm to verify the Wheeler-DeWitt equation gives ground state energy $E_0= 0.10669043206316431\pm 0.02731991298884283$ and the classical eigensolver algorithm yields $6.159038648121874e\times10^-5$ taking cosmological constant $\Lambda=0.581$. The optimized variational circuit is shown in Fig. \ref{qgc_Fig5}.

\subsection{Kaluza-Klein Theory}
Kaluza-Klein theory \cite{qgc_BeciuINC1985} attempts to unify gravity with electromagnetism by allowing for a 5th dimension beyond the usual four dimensions of spacetime. It is considered as a precursor to string theory.The 4-dimensional part of the model can be taken as a Friedmann closed space. The extra dimension y is supposed to be disconnected from the rest so that the y dependence can be factorized out of the metric components. So,

\begin{equation}
ds^2=\lambda^2(y)(-N^2dt^2+a^2(t)d\Omega^2+\phi^2(t)dy^2)
\end{equation}

where $d\Omega^2$ is the metric on a unit 3-sphere, and $phi$ is a scalar field. The 5-dimensional Einstein field equations yield the 4-dimensional Einstein field equations, Maxwell's electromagnetic field equations and an equation for the scalar field. Taking $q_1=a(a+\phi)/2$ and $q_2=a(a-\phi)/2$, we get the Hamiltonian:

\begin{equation}
H_4=\frac{N}{24\pi^2}[k(q_1-q_2)+(p_1^2-p_2^2)]
\end{equation}

On applying the VQE algorithm for this Hamiltonian taking $k=10^{-6}$, we get the ground state eigenvalue $E_0=-0.03330932911207404 \pm 0.19977149262935648$. The classical exact eigensolver algorithm gives the eigenvalue as -1.5714280000014587. The optimized quantum circuit representing the universal wavefunction is shown in Fig. \ref{qgc_Fig6}.

\section{Quantum Circuits \label{qgc_Sec7}}
\begin{figure}[H]
\centering
\includegraphics[width=1\linewidth,height=18cm]{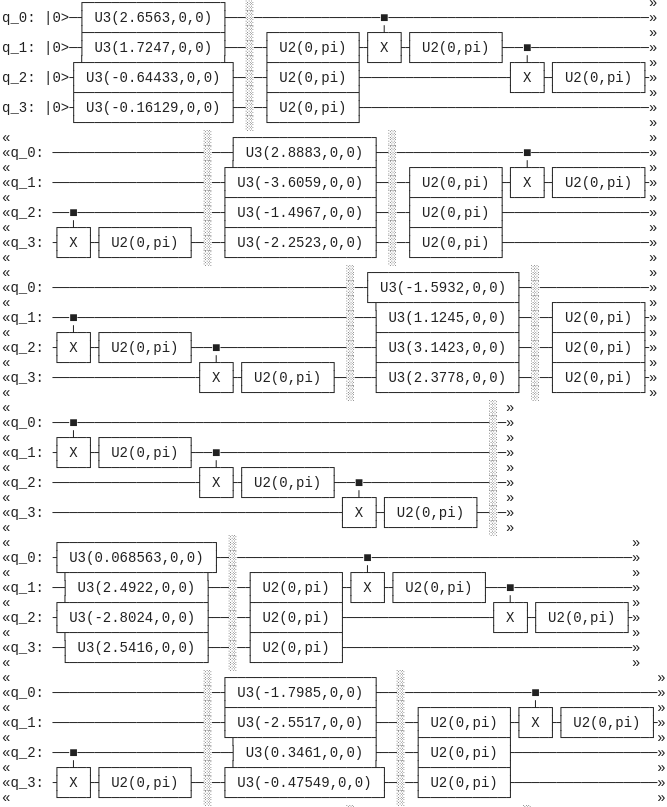}
\includegraphics[width=1\linewidth,height=3cm]{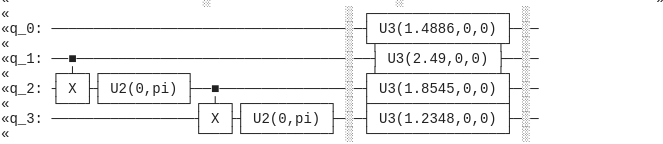}
\caption{Optimized quantum circuit generating the state $\psi$ for Hamiltonian $H_1$}
\label{qgc_Fig3}
\end{figure}

\begin{figure}[H]
\centering
\includegraphics[width=1\linewidth,height=18cm]{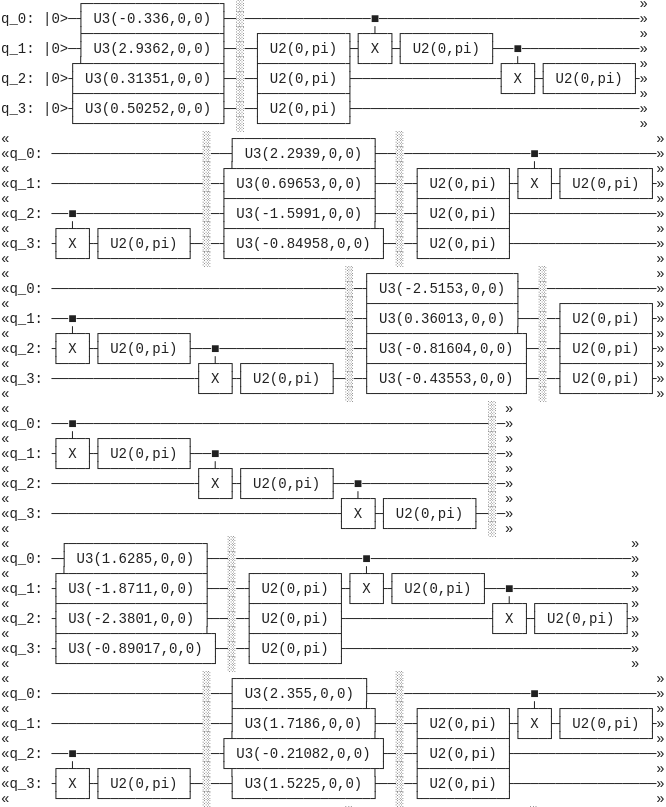}
\includegraphics[width=1\linewidth,height=3cm]{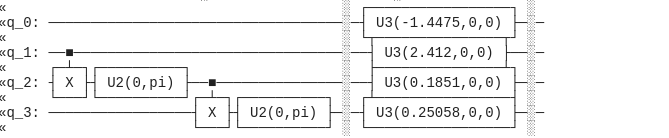}
\caption{Optimized quantum circuit generating the state $\psi$ for Hamiltonian $H_2$}
\label{qgc_Fig4}
\end{figure}

\begin{figure}[H]
\centering
\includegraphics[width=1\linewidth,height=18cm]{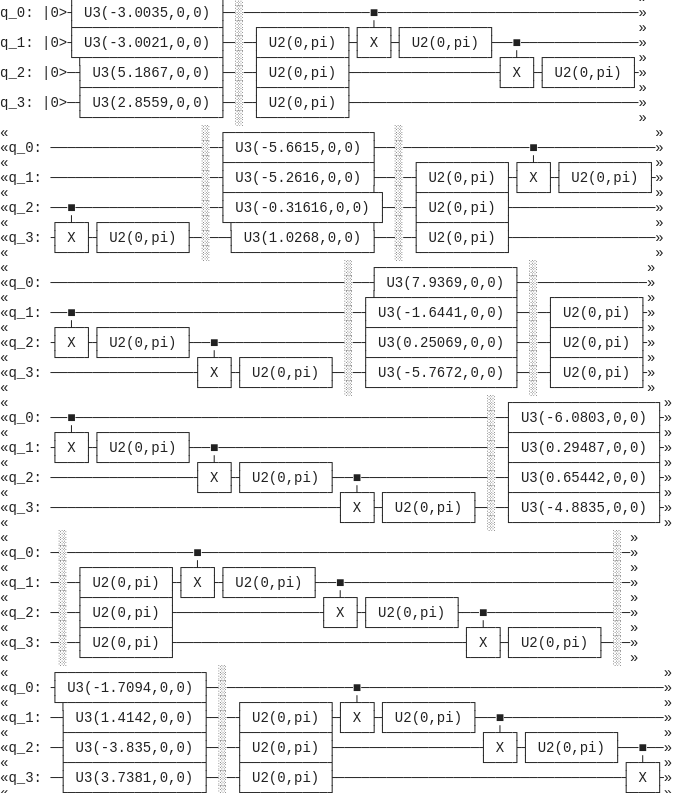}
\includegraphics[width=1\linewidth,height=3cm]{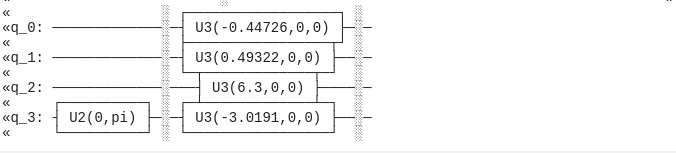}
\caption{Optimized quantum circuit generating the state $\psi$ for Hamiltonian $H_3$}
\label{qgc_Fig5}
\end{figure}

\begin{figure}[H]
\centering
\includegraphics[width=1\linewidth,height=18cm]{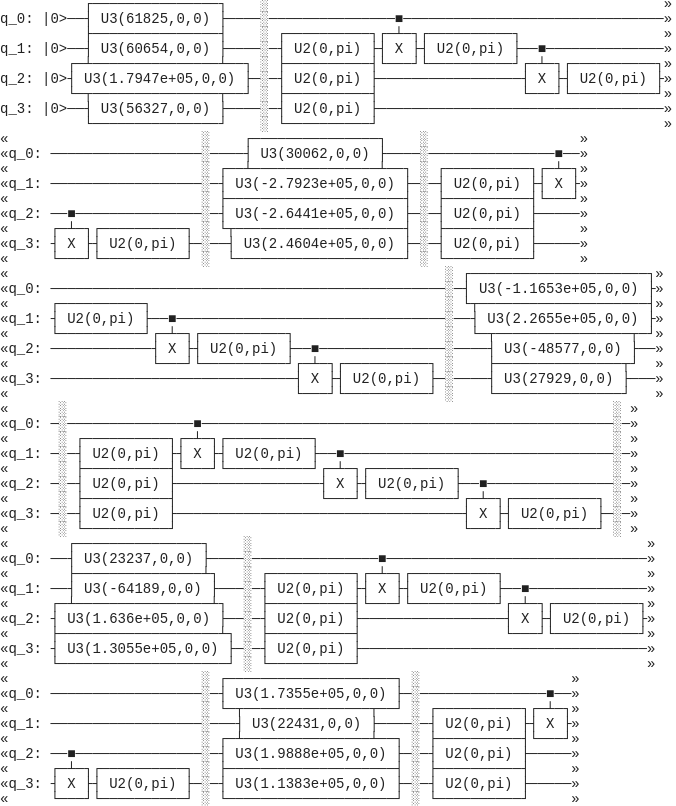}
\includegraphics[width=1\linewidth,height=3cm]{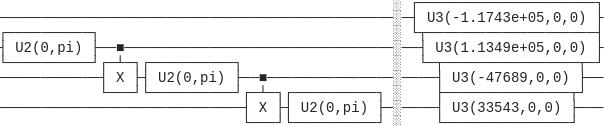}
\caption{Optimized quantum circuit generating the state $\psi$ for Hamiltonian $H_4$}
\label{qgc_Fig6}
\end{figure}

We present the optimized quantum circuits that generate the universal wavefunction for each of the models taken. The initial state for the VQE algorithm was taken as the Ry variational form provided by the IBM QISKit library. The circuit in Fig. \ref{qgc_Fig3} generates the universal wavefunction for Hamiltonian $H_1$ representing a Bianchi-IX type universe with classical radiation density constant $\gamma$, cosmological constant $\Lambda$ and small anisotropies. Fig. \ref{qgc_Fig4} shows the circuit for generating the statevector corresponding to the Hamiltonian $H_2$ obtained for our higher derivative cosmological model. Fig. \ref{qgc_Fig5} shows the circuit generating the universal wavefunction for Hamiltonian $H_3$ of the string dilaton cosmological model taken. The circuit in Fig. \ref{qgc_Fig6} is the representation of $H_4$ corresponding to a Kaluza-Klein model of quantum cosmology.

\section{Conclusion}\label{qgc_Sec8}
It has been previously shown how quantum computing is an efficient tool for studying the homogeneous isotropic FRW minisuperspace with conformally coupled scalar fields \cite{qgc_KocherIEEE2018}. Here, we have successfully utilized the VQE algorithm for finding the universal wavefunction solutions consistent with the Wheeler De Witt constraint applied on anisotropic minisuperspaces, Kaluza-Klein models, higher derivative quantum cosmological models, and string dilaton cosmology. It yielded the upper bounds on the ground state energy and the optimized quantum circuits that generate the required state vectors. The variational method has thus proved beneficial in circumventing the problematic necessity of boundary conditions \cite{qgc_VilenkinPRD2018,qgc_HartlePRD1987} to study the universal wavefunction and can be used in the future for studying other cosmological models or even extending minisuperspace quantum cosmology to the actual superspace, in which the wavefunction exists. The simulation was done using 2 qubits for each dimension of all the 2-dimensional systems considered. The backend used was the ``state vector simulator" provided by QISKit.

\section*{Acknowledgments}
\label{qlock_acknowledgments}
A.G. would like to thank IISER Kolkata for providing hospitality during the course of the project work.A.G and B.K.B. acknowledge the support of Institute fellowship provided by IISER Kolkata. The authors acknowledge the support of IBM Qiskit Python library. The views expressed are those of the authors and do not reflect the official policy or position of IBM or the IBM Qiskit team. A.G. would like to thank Pratyusha Chowdhury \cite{qgc_PratyushaIITG} for useful discussions regarding the paper.

\end{document}